\documentstyle[prd,aps,eqsecnum]{revtex}
\newcommand{\bmit}[1]{\mbox{\boldmath $#1$}}
\newcommand{\alf}{\left[(1-2\mu)^2 -4 \mu^2 e^2\right]}
\newcommand{\alfs}{\left(1-2\frac{M}{a_0}\right)}
\newcommand{\bet}{1-\mu (3+e^2)}

\newcommand{\g}{\left[(1-6\mu)^2 - 4 \mu^2 e^2\right]}
\newcommand{\gs}{\left(1-6\,\frac{M}{a_0}\right)^{\frac{5}{2}}}
\newcommand{\gut}{\left[1-8\mu+4\mu^2 (3+e^2)\right]}
\newcommand{\guq}{\left[1-2\mu (3+e^2)\right]}
\newcommand{\gdq}{\left[1-\mu (7+e^2)\right]}
\newcommand{\gdqs}{\left(1-7\,\frac{M}{a_0}\right)}
\begin{document}

\draft
\preprint{Submitted to Classical and Quantum Gravity}
\title{Lagrangian planetary equations in Schwarzschild space--time}
\author{Mirco Calura, Enrico Montanari\footnote{Electronic address: 
montanari@fe.infn.it}, and P. Fortini}
\address{Department of Physics, University of Ferrara and
INFN Sezione di Ferrara, Via Paradiso 12,
I-44100 Ferrara, Italy}
\maketitle
\begin{abstract}
We have developed a method to study the effects of a perturbation to 
the motion of a test point--like object in a Schwarzschild 
space--time. Such a method is the extension 
of the Lagrangian 
planetary equations of classical celestial mechanics
into the framework of the 
full theory of general relativity. The method provides a 
natural approach to account for relativistic effects in the 
unperturbed problem in an exact way.  
\end{abstract}

\section{Introduction}

The problem of a binary system bound by gravitational interaction
and undergoing the influence of an external force has been 
widely studied within the framework of classical celestial 
mechanics~\cite{mcc63,bro61,roy65,boc96}. 
A possible approach to the problem, taking into account first 
post--Newtonian effects, has been recently proposed~\cite{cal97}.
In this paper we focus our attention to a binary system one body of 
which is much more massive than the other one and achieve the 
planetary equations, describing the time variation of orbital elements
induced by an external perturbation. This is accomplished by starting 
from the exact solution to the equations of motion for a test particle 
in Schwarzschild space--time~\cite{cha83}. Physically the problem to 
be solved is the perturbation of the motion of a point--like object of 
mass $m$ around a non--rotating one of mass $M\gg m$. Since 
relativistic effects have already been considered in the unperturbed 
problem, the only quantity that is assumed to be small is the strength 
of the external force. This offers significant advantage over a 
semiclassical approach, which would have otherwise considered 
relativistic terms as a perturbation.

The paper is organized as follows. In Sec. II we provide a review of 
the exact solution to time--like geodesic equations in Schwarzschild 
space--time. In Sec. III we achieve planetary equations for a generic 
external force. Finally in Sec. IV we provide two applications to show
the capability of the method. In the first case the perturbation is a 
drag force due to interstellar dust while the second case concerns 
with an interstellar magnetic field, in which the less massive body is 
assumed to be charged. However the procedure provides a method that 
can be useful to a general situation.

\section{Unperturbed time--like geodesics}

Let us consider two point--like spinless bodies whose masses are $m$ 
and $M$; 
if $m$ is negligible with respect to $M$, then the motion is given 
by the following Lagrangian (e.g.~\cite{mtw,l2}):
\begin{equation}
{\cal L} = -m\,\sqrt{-\left(\frac{ds}{dt}\right)^2} =
-m\,\frac{d\tau}{dt},
\label{lagrangiana}
\end{equation}
where $d\tau^2 = - ds^2$, and
\begin{equation}
ds^2 = -\left(1 - \frac{2M}{r}\right)\,dt^2 +
\frac{dr^2}{\left(1 - \frac{2M}{r}\right)} + r^2\,\left(
d\theta^2 + \sin^2{\theta}\ d\varphi^2\right)
\label{metrica}
\end{equation}
is the Schwarzschild line element,
in which we have assumed $M$ to be at rest on the origin (hereinafter
$c=G=1$; conventions and notations as in Ref.~\cite{mtw}).
Motion equations are derived from Lagrangian~(\ref{lagrangiana}) in 
the usual way:
\begin{equation}
\frac{d}{dt}\left(\frac{\partial {\cal L}}{\partial \bmit v}\right) 
- \frac{\partial {\cal L}}
{\partial \bmit{x}} = 0,
\label{eqimp}
\end{equation}
where $\bmit x$ and $\bmit v$ are the radius vector 
and velocity of $m$ respectively. Assuming vector 
$\bmit x$ (and $\bmit v$) to be
expressed in cartesian coordinates, the solution to motion equations 
can be written as:
\begin{eqnarray}
x^1 &=& \left(\cos{\omega}\, \cos{\Omega} -
\sin{\omega}\, \sin{\Omega}\, \cos{i}\right)\, \tilde{x}^1 -
\left(\sin{\omega}\, \cos{\Omega} +
\cos{\omega}\, \sin{\Omega}\, \cos{i}\right)\, \tilde{x}^2,
\nonumber\\
x^2 &=& \left(\cos{\omega}\, \sin{\Omega} +
\sin{\omega}\, \cos{\Omega}\, \cos{i}\right)\, \tilde{x}^1 -
\left(\sin{\omega}\, \sin{\Omega} -
\cos{\omega}\, \cos{\Omega}\, \cos{i}\right)\, \tilde{x}^2,
\nonumber\\
x^3 &=& \sin{\omega}\, \sin{i}\  \tilde{x}^1 +
\cos{\omega}\, \sin{i}\  \tilde{x}^2,
\label{eulero}
\end{eqnarray}
where $\omega$, $\Omega$ and $i$ are the usual Euler 
angles, defining the rotation that connects the 
observation reference frame with the intrinsic frame of the 
motion. In classical celestial mechanics they are usually referred 
to as {\em argument of periastron} (the angle in orbital plane 
from the line of nodes to the perihelion point), {\em longitude of 
the ascending node} (the angle measured from the positive $x$ axis 
of the observer to the line of nodes) and {\em inclination of the 
orbit} (the angle between the orbital plane and the $x$--$y$ plane 
of the observer), respectively~\cite{gold80}. Besides 
\begin{eqnarray}
\tilde{x}_1&=&r\,\cos{\phi},\\
\tilde{x}_2&=&r\,\sin{\phi}
\end{eqnarray}
is the solution to the problem in the particular reference frame 
whose $\tilde x^3$ axis is normal to the plane of the orbit. In the
parameterization of Ref.~\cite{cha83} one has:
\begin{eqnarray}
r &=& \frac{a(1-e^2)}{1+ e \cos{\chi}},\qquad\qquad
0\leq e < 1,
\label{r} \\
\phi &=& \frac{2}{\sqrt{1-6\mu +2 \mu e}}\,
\left[F(\frac{\pi}{2}-\frac{\chi}{2}) - 
F(\frac{\pi}{2})\right],
\label{fini}\\
F(\Psi) &=& \int_{0}^{\Psi}\,\frac{d\lambda}
{\sqrt{1-k^2\,\sin^2{\lambda}}}, \\ 
\mu &=& \frac{M}{a\,(1-e^2)},\qquad
k^2 = \frac{4\mu e}{1-6\mu + 2\mu e},\\
t-T &=& -\frac{E}{L}\,\int_{0}^{\chi}\,\frac{d\phi}{d\chi}\,
\frac{r^2}{(1-\frac{2M}{r})}\,d\chi,
\label{t}\\
L&=&\frac{\sqrt{a(1-e^2)M}}{\sqrt{1-\mu\,(3+e^2)}},\label{LL}\\
E&=&\frac{L}{\sqrt{a(1-e^2)M}}\,
\sqrt{(2\mu -1 )^2 - 4 \mu^2 e^2},\label{En}\\
\frac{dt}{d\tau} &=& \frac{E}{1-\frac{2\,M}{r}}. \label{dtdtau}
\end{eqnarray}
The other elements of the orbit $a$, $e$, and $T$ are the relativistic 
extension of the usual Keplerian parameters, which they reduce to in 
the classical limit $\mu=0$, namely the {\em semimajor axis} 
of the ellipse, the {\em eccentricity}, and the {\em time of periastron} 
passage respectively. 
In Eqs.~(\ref{fini}) and~(\ref{t}), integration constants were 
chosen so that $\phi=0$ and $t=T$ for $\chi=0$.

Furthermore, in order to allow the orbits to be confined within 
$r_1 \leq r \leq r_2$~\cite{cha83}, the following inequality must 
hold true
\begin{equation}
\mu \leq \frac{1}{2\,\left(3+e\right)}.
\label{disugmu}
\end{equation}
The orbits described through 
parameterization~(\ref{r})--(\ref{dtdtau}),
together with inequality~(\ref{disugmu}), are the relativistic 
analogues of usual Keplerian ones, to which they reduce in the 
limit $\mu \rightarrow 0$. If condition~(\ref{disugmu}) is not met, 
there do not 
exist stable orbits and the body will eventually plunge into the 
singularity~\cite{cha83}.

\section{Relativistic planetary equations}

The solution considered so far holds true under the assumptions that
the motion of $M$ is negligible and no external force is 
involved. When either of these perturbations cannot be neglected, the
problem can still be approached using Lagrangian~(\ref{lagrangiana}) 
provided that a perturbation term $\bmit Q$ is added to 
the right--hand side of Eq.~(\ref{eqimp})~\cite{gold80}. Namely:
\begin{equation}
\frac{d}{dt}\left(\frac{\partial {\cal L}}{\partial \bmit v}\right) 
- \frac{\partial {\cal L}}
{\partial \bmit{x}} = \bmit Q.
\label{eqpert}
\end{equation}
If $\bmit Q$ may be considered to be small, one is allowed to use a 
perturbative approach to the problem. To this aim we developed a 
procedure that is the general relativistic analogue of the usual 
Lagrangian planetary equations, in a Schwarzschild 
space--time~\cite{mcc63,bro61,roy65,boc96}.
We start by defining the Hamiltonian in the usual way:
\begin{equation}
{\cal H} = \bmit p \frac{d \bmit x}{dt} - {\cal L},\qquad\qquad
\bmit p = \frac{\partial {\cal L}}
{\partial \left(\frac{d\bmit x}{dt}\right)}.
\label{hamiltoniana}
\end{equation}
Hamilton equations are thus written as:
\begin{equation}
\frac{d\bmit x}{dt} = 
\frac{\partial {\cal H}}{\partial \bmit p},\qquad\qquad
\frac{d\bmit p}{dt} = 
- \frac{\partial {\cal H}}{\partial \bmit x} + \bmit Q.
\label{hameqs}
\end{equation}
In order to find out planetary equations one requires that the 
solution to above equations has the same form of 
Eqs.~(\ref{eulero})--(\ref{dtdtau}); 
obviously this implies that the orbital parameters $a$, $e$, $T$, 
$\omega$, $\Omega$, and $i$ vary with time. Therefore, setting
$\bmit x=\bmit x(C_j,t)$, where $C_j$ ($1 \le j \le 6$) are the orbital 
elements $a,\,e,\,T,\,\omega,\,\Omega,\,i$, we have:
\begin{equation}
\frac{\partial\bmit x}{\partial t} = 
\frac{\partial {\cal H}}{\partial \bmit p},\qquad\qquad
\frac{\partial \bmit p}{\partial t} = 
- \frac{\partial {\cal H}}{\partial \bmit x},
\label{hameqsimp}
\end{equation}
where now
\begin{equation}
\bmit p = \frac{\partial {\cal L}}{\partial 
\frac{\partial\bmit x}{\partial t}}
\label{pimp}
\end{equation}
and $\frac{d C_j}{dt}$ are defined in such a way that
$\bmit v = \frac{d\bmit x}{dt} = \frac{\partial\bmit x}{\partial t}$; 
this is accomplished if:
\begin{equation}
\sum_{j=1}^{6} 
\frac{\partial \bmit x}{\partial C_j}\,\frac{d C_j}{dt} = 0,
\label{cjcond}
\end{equation}
as developed in Ref.~\cite{mcc63}.
Using the definition for $\bmit p$ of Eq.~(\ref{pimp}), together with
Eqs.~(\ref{hameqsimp}) and~(\ref{cjcond}), Hamilton 
equations~(\ref{hameqs}) can be rewritten introducing the so--called 
Lagrangian brackets~\cite{gold80}:
\begin{equation}
\left[C_j,\, C_k\right] \stackrel{def}=
\frac{\partial\bmit{x}}{\partial C_j}\cdot
\frac{\partial\bmit{p}}{\partial C_k} -
\frac{\partial\bmit{x}}{\partial C_k}\cdot
\frac{\partial\bmit{p}}{\partial C_j}.
\label{defparentesi}
\end{equation}
This way 
the equations describing the 
time evolution of the orbital elements (otherwise constant) induced by
the external perturbation are:
\begin{equation}
\sum_{k=1}^{6}\,[C_j,C_k]\,\frac{dC_k}{dt} = {\cal F}_j,\qquad\qquad
{\cal F}_j = \frac{\partial\bmit x}{\partial C_j} \cdot \bmit{Q}.
\label{system}
\end{equation}
Therefore the problem is reduced to the solution of the above linear 
system in the variables $\frac{dC_j}{dt}$. 
The convenience of this approach owes to the independence of
Lagrangian brackets upon time explicitly, as one can see 
through Eqs.~(\ref{hameqsimp}). The classical version of this 
property is extensively treated in Ref.~\cite{mcc63}. This way we can 
calculate them at convenient time, such as $t=T$~\cite{mcc63,cal97}.
After tedious but straightforward computation we obtain the only 
non--vanishing brackets:
\begin{eqnarray}
\left[a,\,T\right]&=&\frac{m\,M}{2\,a^2}\,
\frac{\left[1-8\mu+4\mu^2\,(3 + e^2)\right]}
{\sqrt{(1-2\mu)^2 - 4\mu^2 e^2}\,\left[1-\mu \,(3+e^2)
\right]^{\frac{3}{2}}},\\
\left[e,\,T\right]&=&\frac{4\,m\,e(1-e^2)\,\mu^3}
{\sqrt{(1-2\mu)^2 - 4\mu^2 e^2}\,\left[1-\mu \,(3+e^2)
\right]^{\frac{3}{2}}}, \label{eT}\\
\left[\Omega,\,i\right]&=&-\,m\,
\sqrt{M\,a\,(1-e^2)}\,\frac{\sin{i}}{\sqrt{1-\mu\,(3+e^2)}},\\
\left[a,\,\omega\right]&=&-\frac{m}{2}\,
\sqrt{\frac{M\,(1-e^2)}{a}}\,\frac{1-2\mu\,(3+e^2)}
{\left[1-\mu (3+e^2)\right]^{\frac{3}{2}}},\\
\left[a,\,\Omega\right]&=&\left[a,\,\omega\right]\,\cos{i},\\
\left[e,\,\omega\right]&=&m\,e\,\sqrt{\frac{M\,a}{1-e^2}}\,
\frac{1-\mu (7+e^2)}{\left[1-\mu (3+e^2)
\right]^{\frac{3}{2}}},\\
\left[e,\,\Omega\right]&=&\left[e,\,\omega\right]\,\cos{i}.
\end{eqnarray}
We notice that, except for Eq.~(\ref{eT}), the leading term in the 
above brackets is the classical one, in the limit $\mu \rightarrow 0$.
As for $[e,T]$, its classical limit vanishes.
We are now in a position to invert system~(\ref{system}) and obtain 
the time evolution of orbital elements:
\begin{eqnarray}
\frac{da}{dt}&=&-\,\frac{2a\,
\sqrt{\alf\,\left[\bet\right]\,\gdq}}{m\,(1-e^2)\mu\g}\,{\cal F}_3 \nonumber \\ 
&+& \frac{8\,\sqrt{\mu^3}\sqrt{\bet}}{m\,\g}\,{\cal F}_4, \label{dadt}\\
\frac{de}{dt}&=&-\,\frac{\sqrt{\alf\,\left[\bet\right]}\,\guq}
{\mu\,m\, e\,\g}\,{\cal F}_3\nonumber\\ 
&-&\frac{\sqrt{\bet}\,\gut}{m\,e\,\sqrt{\mu}\,\g a}\,{\cal F}_4,
\label{dedt}\\
\frac{dT}{dt}&=&\frac{2a\,\sqrt{\alf\,\left[\bet\right]}\,\gdq}
{m\,(1-e^2)\,\mu\,\g}\,{\cal F}_1\nonumber\\ 
&+&\frac{\sqrt{\alf\,\left[\bet\right]}\,\guq}{m\,e\,\mu\,\g}\,{\cal F}_2,\\
\frac{d\omega}{dt}&=&-\,\frac{8\,\sqrt{\mu^3}}{m\,\g}\,{\cal F}_1 +
\frac{\sqrt{\bet}\,\gut}{m\,e\,\sqrt{\mu}\,\g\,a}\,{\cal F}_2\nonumber\\ 
&-&\frac{\sqrt{\bet}\,\cot{i}}{m\,\sqrt{\mu}\,a(1-e^2)}\,{\cal F}_6,\\
\frac{d\Omega}{dt}&=&\frac{\sqrt{\bet}}{m\,\sqrt{\mu}\,a(1-e^2)\,
\sin{i}}\,{\cal F}_6,\label{dOdt}\\
\frac{d\cos{i}}{dt}&=&-\,\frac{\sqrt{\bet}\cos{i}}{m\,\sqrt{\mu}\,a(1-e^2)}
\,{\cal F}_4 +
\frac{\sqrt{\bet}}{m\,\sqrt{\mu}\,a(1-e^2)}\,
{\cal F}_5. \label{didt}
\end{eqnarray}
For any assigned external perturbation, it is possible to calculate 
its effects on orbital elements by solving 
system~(\ref{dadt})--(\ref{didt}); as in classical celestial mechanics 
the first--order time variation of orbital elements is accomplished by
substituting the unperturbed values into the right--hand side of 
Eqs.~(\ref{dadt})--(\ref{didt}).

\section{Applications}

In this Section we provide two applications of the proposed method.
In the former we consider a dissipative effect, which has a classical 
analogue. In the latter we show an effect which is merely 
relativistic.

\subsection{Drag force}

As a first application we consider the case in which the external 
perturbation is drag due to dust.
Let us suppose to be within the range of velocities for which the drag 
force can be written as~\cite{mcc63}:
\begin{equation}
\bmit F=-f\,|\bmit v|\,\bmit v,\qquad\qquad
f = \frac{C_D\,{\cal S}\,\rho}{2},
\label{dragforce}
\end{equation}
where $C_D$ is the {\em drag coefficient}, ${\cal S}$ is the
{\em cross--sectional area} of the body, and $\rho$ is the {\em dust 
density}, which we assume to be constant. Comparing 
Eq.~(\ref{dragforce}) with Eqs.~(\ref{eqpert}) and~(\ref{dtdtau}) we 
achieve:
\begin{equation}
\bmit Q=- \frac{f\,E\,|\bmit v|\,\bmit v}
{\left(1-\frac{2\,M}{r}\right)^2}.
\end{equation}
When this force is substituted in the last of Eqs.~(\ref{system})
we get
\begin{eqnarray}
{\cal F}_j&=&-\frac{f\,E\,|\bmit v|}{\left(1-\frac{2\,M}{r}\right)^2}
\,\left(\frac{\partial \tilde{x}_1}
{\partial t}\,
\frac{\partial \tilde{x}_1}{\partial C_j} + 
\frac{\partial \tilde{x}_2}{\partial t}
\,\frac{\partial \tilde{x}_2}{\partial C_j}\right),
\qquad j=1,2,3,\\
{\cal F}_4&=&\frac{f\,E\,|\bmit v|}{\left(1-\frac{2\,M}{r}\right)^2}
\,\left(
\frac{\partial \tilde{x}_1}{\partial t}\,\tilde{x}_2 -
\frac{\partial \tilde{x}_2}{\partial t}\,\tilde{x}_1\right),\\
{\cal F}_5&=&{\cal F}_4\,\cos{i},\label{f5}\\
{\cal F}_6&=&0.\label{f6}
\end{eqnarray}
The first result we obtain is that from Eqs.~(\ref{f5}) and~(\ref{f6}) 
$i$ and $\Omega$ keep constant even in the perturbed case.
The main feature of the perturbation is a secular 
decreasing of $a$ which results in a spiraling of $m$ around $M$.
Since we are only interested in secular variation we perform an 
average on Eq.~(\ref{dadt}) over $2\,\pi$ in $\chi$. For the classical 
approach to this problem see~\cite{mcc63}. To provide an analytically 
simple example,
we assume the parameter $e$ to be small. Up to the first order in the 
expansion in power of $e$ we get:
\begin{eqnarray}
\left<\frac{da}{d\chi}\right> &=& \frac{2\,a_0^2\,f\,\gdqs}{m\,\gs}\,
\left[ 1- \frac{9\,M\,e_0}{a_0\,\left(1-6\,
\frac{M}{a_0}\right)}\right] \nonumber\\ 
&+&\frac{8\,M^2\,f}{m\,\alfs\,\gs}\,
\left[ 1- \frac{6\,M\,e_0}{a_0\,\left(1-6\,
\frac{M}{a_0}\right)}\right],
\end{eqnarray}
where we have replaced
the orbital elements in the right hand side of Eq.~(\ref{dadt}) with 
the initial ones, up to the first order in the strength of $Q$.
It is interesting to notice that when the body reaches $a=6\,M$ in the 
motion, the parameterization used for the solution breaks 
down [see Eq.~(\ref{disugmu})], regardless the value of parameter $e$,  
and the body would plunge into the singularity even if there were no 
dust~\cite{cha83}.

This method could also be used to determine the variation of 
orbital elements caused by gravitational wave emission by the
lightest body. To this
aim it suffices to find the form of the back--reaction force due to 
the power loss. This force could be obtained starting from the energy 
emitted~\cite{l2,poi93,cut93}.

\subsection{External magnetic field}

As a second example we study the perturbation 
to the motion of an electric charge $q$ with mass $m$
induced by an external magnetic field.
We consider a particular magnetic field which is constant and 
homogeneous at infinity and directed along $x^3$ axis.
The electromagnetic tensor--solution 
to Maxwell equations in Schwarzschild space-time with the above 
boundary condition--is herewith reported in terms of its non--vanishing 
contravariant cartesian components:
\begin{eqnarray}
F^{13}&=&-\,F^{31}=-\,\frac{2\,M{\cal B}}{r^3}\,x^2\,x^3,\\
F^{23}&=&-\,F^{32}=   \frac{2\,M{\cal B}}{r^3}\,x^1\,x^3,\\
F^{12}&=&-\,F^{21}= {\cal B} - \frac{2\,M{\cal B}}{r^3}\,
\left[(x^1)^2 + (x^2)^2\right],
\label{campi}
\end{eqnarray}

The motion of a charged particle in an external electromagnetic field 
is described by~\cite{mtw}
\begin{equation}
m\,\left(\frac{d^2 x^\mu}{d\tau^2} + 
\Gamma^\mu_{\alpha\beta}\,\frac{dx^\alpha}{d\tau}\,
\frac{dx^\beta}{d\tau}\right) = 
q\,F^\mu_{\ \nu}\,\frac{dx^\nu}{d\tau}.
\label{lorentz}
\end{equation}
The only non--vanishing mixed components of the electromagnetic
tensor are:
\begin{eqnarray}
F^1_{\ 1} &=& - F^2_{\ 2} = \frac{2\,M\,{\cal B}}{r^3}\,x^1\,x^2, \\
F^1_{\ 2} &=& {\cal B} - \frac{2\,M\,{\cal B}}{r^3}\,(x^1)^2, 
\qquad\qquad
F^2_{\ 1} = - {\cal B} + \frac{2\,M\,{\cal B}}{r^3}\,(x^2)^2, \\
F^3_{\ 1} &=& \frac{2\,M\,{\cal B}}{r^3}\,x^2\,x^3,\qquad\qquad
F^3_{\ 2} = - \frac{2\,M\,{\cal B}}{r^3}\,x^1\,x^3.
\end{eqnarray}
Comparing Eq.~(\ref{lorentz}) with Eqs.~(\ref{eqpert}) 
and~(\ref{dtdtau}) we get:
\begin{equation}
Q^k = \frac{q}{1-\frac{2\,M}{r}}\,F^k_{\ j}\,v^j.
\label{qmagn}
\end{equation}
We also assume that the magnetic field is weak enough to induce only a 
small perturbation to the motion. This assumption allows one to solve
planetary equations~(\ref{dadt})--(\ref{didt}) in a perturbative
way. 

We focus our attention to the orbits lying 
in a plane perpendicular to $x^3$ axis, so we set $i=0$. 
Moreover we set 
$\Omega = 0$. Later on we will prove that orbits satisfying these 
assumptions do exist.  
With this choice $F^3_{\ 1} = F^3_{\ 2} = 0$. 
The non--vanishing components of $\bmit Q$ read:
\begin{eqnarray}
Q^1 &=& \frac{q}{1-\frac{2\,M}{r}}\, \left(
F^1_{\ 1}\,v^1 + F^1_{\ 2}\,v^2 \right),\\
Q^2 &=& \frac{q}{1-\frac{2\,M}{r}}\, \left(
F^2_{\ 1}\,v^1 - F^1_{\ 1}\,v^2 \right).
\end{eqnarray}
Therefore by means of the last of Eqs.~(\ref{system}) and 
assuming $i = 0$ we obtain 
\begin{eqnarray}
{\cal F}_4 &=& {\cal F}_5 = - \frac{q\,{\cal B}}{2\,\left(
1-\frac{2\,M}{r}\right)}\, \frac{\partial\ }{\partial t}\,
\left(r^2\right)\label{f4f5},\\
{\cal F}_6 &=& 0;
\end{eqnarray}
these equations imply that Eq.~(\ref{didt}) 
is identically satisfied and $\Omega$ is a constant we 
can take equal to zero. Besides, we get
\begin{equation}
{\cal F}_3 = \frac{q\,L}{E}\,\frac{M\,{\cal B}}{r^3}\, 
\frac{\partial\ }{\partial t}\,\left(r^2\right).
\label{f3}
\end{equation}
This way we see from Eq.~(\ref{dadt}) that, 
as opposite to the classical case where ${\cal F}_3=0$ and the 
coefficient of ${\cal F}_4$ is also zero, $a$ is not a constant of 
the motion. If $t$ is replaced by $\chi$ Eq.~(\ref{dadt}) becomes
the following equation in 
the first perturbation order (keeping orbital elements constant 
in the right hand side):
\begin{eqnarray}
\frac{da}{d\chi} &=& \frac{4\,q\,{\cal B}}{m}\,\sqrt{a^5\,M\,(1-e^2)}
\,\frac{\sqrt{\left[\bet\right]\,\gdq}}{\g}\,\frac{d\ }{d\chi}\frac{1}{r} 
\label{dadchi}\\
&-& \frac{8\,q\,{\cal B}}{m}\,\sqrt{\frac{M^3}{a^3\,(1-e^2)^3}}\,
\frac{\sqrt{\bet}}{\g}\,\frac{d\ }{d\chi}\left[
\frac{r^2}{2} + 2\,M\,r + 4\,M^2\,
\log{\frac{r\left(1-\frac{2\,M}{r}\right)}
{r_0\left(1-\frac{2\,M}{r_0}\right)}}\right].\nonumber
\end{eqnarray}
From the above expression it is straightforward there is no secular 
variation of $a$: in fact an average of the above equation over 
$2\,\pi$ in $\chi$ is zero since $r$ is a $2\pi$--periodic function of $\chi$
[see Eq.~(\ref{r})].
Last equation can be solved to obtain the explicit dependence 
of $a$ upon $\chi$:
\begin{eqnarray}
a(\chi) &=& a_0 + \frac{4\,e\,q\,{\cal B}}{m}\,\sqrt{\frac{a_0^3\,M}{1-e^2}}\,
\frac{\sqrt{\left[\bet\right]\,\gdq}}{\g}\,(\cos{\chi}-\cos{\chi_0}) 
\label{achi}\\
&+& \frac{8\,q\,{\cal B}}{m}\,\sqrt{\frac{M^3}{a_0^3\,(1-e^2)^3}}\,
\frac{\sqrt{\bet}}{\g}\,\left\{
\frac{e\,a_0^2\,(1-e^2)^2\,(\cos{\chi}-\cos{\chi_0})}
{(1+e\,\cos{\chi})^2\,(1+e\,\cos{\chi_0})^2}\times\right.\nonumber\\ 
&&\left.\left[1+\frac{e}{2}\,(\cos{\chi}+\cos{\chi_0})\right]\right.
+ \frac{2\,e\,a_0\,M\,(1-e^2)\,(\cos{\chi}-\cos{\chi_0})}
{(1+e\,\cos{\chi})\,(1+e\,\cos{\chi_0})}\nonumber \\
&+& \left. 4\,M^2\,\log{\frac{(1+e\,\cos{\chi})\,
\left[a_0\,(1-e^2) - 2\,M\,(1+e\,\cos{\chi_0})\right]}
{(1+e\,\cos{\chi_0})\,
\left[a_0\,(1-e^2) - 2\,M\,(1+e\,\cos{\chi})\right]}}\right\}, 
\nonumber
\end{eqnarray}
where $a_0 = a(\chi_0)$. The above expression together with 
Eq.~(\ref{t}) allows one to evaluate $a(t)$.
As one can check through Eqs.~(\ref{dedt}), (\ref{f3}), 
and~(\ref{f4f5}), $e$ does not have secular terms either. This implies 
that Eq.~(\ref{achi}) holds true at any time and hence $r$ ranges 
within an interval.

\section{Conclusion and discussion}

We have derived Lagrangian planetary equations to describe the effect 
of a perturbation to time--like geodesic in a Schwarzschild 
space--time. 
Our method provides a natural way to study the evolution of binary 
systems, when relativistic effects can not be neglected, 
as for instance when coalescing stage is approached.

The results we have obtained in this paper hold true for a test 
particle; from the physics point of view this means that the effect 
of the mass of the orbiting particle gives
rise to a perturbation term $\bmit Q_m$ that can be neglected with 
respect to the one, $\bmit Q_{ext}$, arising from an external 
perturbation. The order of magnitude of the former is expected to be 
the classical one, that is $\bmit Q_m \sim m^2/r^2$. If the condition 
$\bmit Q_m \ll \bmit Q_{ext}$ ceases to hold, the problem deserves a 
further investigation.

Other than the two examples considered
our method can also be useful to other kinds of 
perturbations such as 
oblateness or rotation of the orbiting star, energy loss caused by 
gravitational wave emission, or interaction with a third body.

\acknowledgments

The authors are pleased to thank V. Guidi for reading of the 
manuscript.


\begin{references}

\bibitem{mcc63} S. W. McCuskey, {\em Introduction to Celestial 
Mechanics} (Addison--Wesley, Reading, MA, 1963)

\bibitem{bro61} D. Brower and G. M. Clemence,
{\em Methods of Celestial Mechanics} (Academic Press, New York, 1961).

\bibitem{roy65} A. E. Roy, {\em The Foundations of Astrodynamics}
(Macmillan, London, 1965).

\bibitem{boc96} D. Boccaletti and G. Pucacco, {\em Theory of Orbits. 
Vol. I: Integrable Systems and Non--perturbative Methods}
(Springer--Verlag, Berlin, Heidelberg, 1996),
{\em Theory of Orbits. Vol. II: Perturbative and Geometrical 
Methods} (Springer--Verlag, Berlin, Heidelberg, 1998) and 
references therein.

\bibitem{cal97} M. Calura, P. Fortini, and E. Montanari, \prd
{\bf 56}, 4782 (1997).

\bibitem{cha83} S. Chandrasekhar, {\em The Mathematical Theory of 
Black Holes} (Oxford University Press, New York, 1983).

\bibitem{mtw} C. Misner, K. S. Thorne and J. A. Wheeler, 
{\em Gravitation} (Freeman, San Francisco, 1973).

\bibitem{l2} L. D. Landau, E. M. Lif\v{s}itz, {\em Course of 
Theoretical Physics Vol. 2: The Classical Theory of Fields} 
(Pergamon Press, 1975).

\bibitem{gold80} H. Goldstein, {\em Classical Mechanics}, 2nd ed. 
(Addison--Wesley, Reading, MA, 1980).

\bibitem{poi93} E. Poisson, \prd {\bf 47}, 1497 (1993). 

\bibitem{cut93} C. Cutler, L. S. Finn, E. Poisson, and G. J. Sussman, 
\prd {\bf 47}, 1511 (1993).

\end{references}
\end{document}